\documentclass[aps,epsf]{revtex4}   

\def\bra{\langle}
\def\ket{\rangle}

\begin{document}

\title{Interrelationship of Isospin and Angular Momentum
}

\author{L. Zamick and A.Z. Mekjian}
\address{Department of Physics, Rutgers University, Piscataway NJ 08854}

\author{S.J. Lee\footnote{ssjlee@khu.ac.kr}}
\address{Department of Physics and Institute of Natural Sciences,
             Kyung Hee University, Suwon, KyungGiDo, Korea}


\begin{abstract}

It is noted that the simple interaction in isospin variables
$a (1/4 - t(i)\cdot t(j))$, in a single $j$ shell calculation,
can also be written with angular momentum variables.
For the configuration $(j^2) J_A$ for even $J_A$ the isospin is one;
for odd $J_A$ it is zero.
Hence the above interaction can also be written as $a (1 - (-1)^{J_A})/2$.
For the $I=0$ state of an even-even Ti isotope with $n$ neutrons,
the hamiltonian matrix element of this interaction is
 $\bra [J'J']_0 |H| [JJ]_0\ket/a = (n+1) \delta_{JJ'}
   - (n+1) \left(j^n Jj|\} j^{n+1} j\right)
           \left(j^n J'j|\} j^{n+1} j\right)$.
The eigenvalues of this interaction can be found by using the
isospin form of the interaction.
They are $(n+1)a$ for $T = |N-Z|/2$ and zero for $T = |N-Z|/2 + 2$.
One can apply this to some extent to obtain the number of pairs of
nucleons with given total angular momentum $J_A$ in a given Ti isotope.



\end{abstract}

\maketitle


\section{Introduction}

In the single $j$ shell there is a simple relationship between the
isospin and angular momentum of a state of the $(j^2)$ configuration.
If the total angular momentum, $J$, is even the isospin $T$ is one;
if $J$ is odd $T$ is zero.
Hence for example for $^{42}$Sc the $f_{7/2}^2$ neutron-proton states
with $J = 0$, 2, 4, and 6 will have isospin $T=1$
while the $J = 1$, 3, 5 and 7 states will have isospin $T = 0$.
Since the later are isosinglet states they cannot occur for a
system of two protons ($^{42}$Ti), or of two neutrons ($^{42}$Ca).
For these two nuclei one gets only $J = 0$, 2, 4 and 6
in the low lying spectrum.
Of course for higher excitation one can get state of odd
angular momentum from other configurations.
One can easily see why the $J = 7$ state cannot occur for two neutrons.
The $J = 7$ $M = 7$ state is $\psi_j^j(1) \psi_j^j(2)$ ($j = f_{7/2}$).
This state is obviously symmetric in space and spin and so cannot
exist for two identical fermions, but it can exist for
a neutron-proton pair.

We can obtain an interesting interrelationship between isospin
and angular momentum by noting that a simple interaction can
be written either with isospin variables or angular momentum variable.
Consider the two particle interaction $V = a (1/4 - t(1)\cdot t(2))$.
This is zero for $T = 1$ states and has a value $a$ for $T = 0$ states.
But we can also write this in terms of the total angular momentum
of the two nucleons $J_A$. $V = a (1 - (-1)^{J_A})/2$.
This vanishes for even $J_A$ but is equal to $a$ for odd $J_A$.
As mentioned in the previous paragraph for even $J_A$, $T=1$
while for odd $J_A$, $T=0$.
The work in this paper is based on this interrelationship
between isospin and angular momentum.

\section{The even-even Ti isotopes}

Consider a Ti isotope with 2 $f_{7/2}$ protons and $n$ $f_{7/2}$ neutrons.
We will use the notation $I$ for the total angular momentum of
a given state of a given Ti isotope and $J_P$ and $J_N$ for the
angular momenta of the protons and neutrons respectively.
For $^{44}$Ti, $^{46}$Ti and $^{48}$Ti 
the values of $n$ are 2, 4 and 6 respectively.
Let us get the Hamiltonian matrix and the eigenvalues with
the simplified interaction $a (1/4 - t(i)\cdot t(j))$. 
We have
\begin{eqnarray}
 \frac{1}{a} \sum_{i<j} V(ij) &=& \frac{1}{4} \frac{A(A-1)}{2}
        - \frac{1}{2} \left(\sum_i t(i)\cdot \sum_j t(j)\right) 
        + \frac{1}{2} \sum_i t_i^2     \nonumber \\
    &=& \frac{A(A-1)}{8} - \frac{1}{2} T(T+1) + \frac{3}{8} A
\end{eqnarray}
where $A$ is the number of valence nucleons $A = n +2$.
We can evaluate this for the $I=0$ states of the Ti isotopes
which have isospin $T = \frac{|N-Z|}{2}$ and $\frac{|N-Z|}{2} + 2$.
\begin{eqnarray}
 \bra V\ket &=& (A-1) a = (n+1) a \hspace{5em} {\rm for \ } 
      T = \frac{|N-Z|}{2} = \frac{n-2}{2}
       \nonumber  \\
 \bra V\ket &=& 0  \hspace{13em} {\rm for \ } 
      T = \frac{|N-Z|}{2} + 2 = \frac{n+2}{2}
           \label{meanv}
\end{eqnarray}
For the lowest isospin states of $^{44}$Ti, $^{46}$Ti and $^{48}$Ti
the values are $3a$, $5a$ and $7a$ respectively.
The reason this interaction vanishes for states with $T = \frac{|N-Z|}{2}+2$
is that these states are double analogs of corresponding states
in the calcium isotopes.
For these `neutron only' systems the interaction can only
take place in $T = 1$ states, but our interaction, by design,
is zero for such states.

We now evaluate the Hamiltonian matrix using angular momentum variables.
For a given Ti isotope the basis states are $[(j^2)_{J_P} (j^n)_{J_N}]_I$
where $I$ is the total angular momentum, $J_P$ is the angular momentum 
of the two protons and $J_N$ of the $n$ neutrons.
If we specialize to $I = 0$ state then $J_P = J_N \equiv J$.

Let us define the two particle interaction matrix element
 $E(J) = \bra (jj) J |V| (jj) J\ket$.
The general expression for the matrix element of Hamiltonian is \cite{ref1}
\begin{eqnarray}
 \bra[J_P' J_N']_I |H| [J_P J_N]_I\ket
     &=& [E(J_P) + E(J_N)] \delta_{J_P J_P'} \delta_{J_N J_N'}   \nonumber  \\
     & & + 2n \sum_{J_B, J_A, J_0} (j^{n-1} J_0 j |\} J_N)
        (j^{n-1} J_0 j |\} J_N')
       \bra(jj) J_P (J_0j) J_N | (jJ_0) J_B (jj) J_A\ket_I  \nonumber \\
          & &   \hspace{0.5in} \times
       \bra(jj) J_P' (J_0j) J_N' | (jJ_0) J_B (jj) J_A\ket_I
       E(J_A)       \label{melemh}
\end{eqnarray}
In the above we have the coefficients of fractional parentage (cfp) 
which isolates one neutron from the others $(n-1)$.
The unitary nine $j$ symbol is needed to combine this neutron with one
of the protons in order to obtain the $n$-$p$ interaction.

We now consider the simplified interaction $a(1-(-1)^{J_A})/2$
and we limit ourselves to $I = 0$ states for which $J_P = J_N = J$
and $J_P' = J_N' = J'$.
With this choice of interaction the $n$-$n$ and $p$-$p$ interactions vanish
and so do the first two terms in Eq.(\ref{melemh}).
We get
\begin{eqnarray}
 \bra[J'J']_{I=0} |H| [JJ]_{I=0}\ket
   &=& 2n \sum_{J_0, J_A} (j^{n-1} J_0 j |\} J) (j^{n-1} J_0 j |\} J')
       \bra(jj)J (J_0j)J | (jJ_0)J_A (jj)J_A\ket_0   \nonumber \\
   & &       \hspace{0.5in} \times
       \bra(jj)J' (J_0j)J' | (jJ_0)J_A (jj)J_A\ket_0
       E(J_A)     \label{mateleh}
\end{eqnarray}
with $E(J_A) = a \frac{(1 - (-1)^{J_A})}{2}$.

The unitary nine $j$ symbol is related to the Wigner $9j$ by
\begin{eqnarray}
  \bra(jj)J (J_0j)J | (jJ_0)J_A (jj)J_A\ket_0
    &=& (2J+1) (2J_A+1)
       \left\{\matrix{j & j & J \cr J_0 & j & J \cr J_A & J_A & 0}\right\}
\end{eqnarray}
We can use the properties of $9j$ symbol to simplify the expression.
We take them from de-Shalit and Talmi \cite{ref6}
and Talmi's 1993 book \cite{ref2}.
We show the formulae in our Appendix A.
We find
\begin{eqnarray}
 \bra[J'J']_0 |H| [JJ]_0\ket /a &=& n \delta_{JJ'}
     - n \sqrt{(2J+1)(2J'+1)}
      \sum_{J_0} (j^{n-1} J_0 j |\} j^n J)  \nonumber \\
          & & \hspace{0.5in} \times
      (j^{n-1} J_0 j |\} j^n J')
      \left\{\matrix{J_0 & j & J' \cr j & j & J}\right\}
               \label{talmi}
\end{eqnarray}
This would seem to be as far as we could go.
However there is a recursion formula involving cfp's 
in Talmi's 1993 book (p.996) which can further simplify the expression
\cite{ref2}
(see also Appendix A)
as long as $j$ is less than or equal to 7/2.
For these $j$ values there is only one $J=0$ state for a system 
of an even number of identical particles in the single $j$ shell.
We get finally
\begin{eqnarray}
 \bra[J'J']_0 |H| [JJ]_0\ket /a &=& (n+1) \delta_{JJ'}
     - (n+1) (j^nJj|\}j^{n+1}j) (j^nJ'j|\}j^{n+1}j)     \label{angl}
\end{eqnarray}
where $n$ is the number of neutrons in a given Ti isotope.

Note that the final expression Eq.(\ref{angl}) yields a separable 
interaction $(j^n J j |\} j^{n+1} j) (j^n J' j |\} j^{n+1} j)$.
For $n = 2$ we get a simple expression, 
from Eqs.(\ref{talmi}) and (\ref{angl}),
\begin{eqnarray}
 3 (j^2 J j |\} j^3 j) (j^2 J' j |\} j^3 j)
   &=& \delta_{JJ'}
      + 2 \sqrt{(2J+1)(2J'+1)} \left\{\matrix{j&j&J'\cr j&j&J}\right\}
         \label{sixj}
\end{eqnarray}
More generally we can get the eigenvalues by comparing Eq.(\ref{angl}) 
with the expression of Eq.(\ref{meanv}) obtained using isospin variables.
For $T = |N-Z|/2$ the expression is $(n+1)a$.
But this is carried by the first term in the angular momentum expression
Eq.(\ref{angl}).
We can thus say that the eigenvalues of the separable interaction
 $(j^n J j |\} j^{n+1} j) (j^n J' j |\} j^{n+1} j)$
are \underline{zero} for states with $T = |N-Z|/2$
and \underline{one} for states with $T = |N-Z|/2 + 2$.
Since the states with $T  = |N-Z|/2$ are all degenerate,
all linear combinations of states with this isospin 
are eigenfunctions of the above interaction.

\section{Application - wave function of the Ti isotopes}

A given wave function of a Ti isotope with total angular
momentum $I$ can be written as
\begin{eqnarray}
 \psi^{\alpha I} = \sum D^{\alpha I}(J_P, J_N v) [(j^2)_{J_P} (j^n)_{J_N}]_I
       \label{psii}
\end{eqnarray}
where $D^{\alpha I}(J_P, J_N v)$ is the probability amplitude that
in a state of total angular momentum $I$
the protons couple to angular momentum $J_P$ and 
the neutrons to $J_N$ with seniority $v$ \cite{ref1}.
The $D$'s form a column vector representation of the wave functions.
The wave functions of the lowest $0^+$ states, 
with isospin $T = |N-Z|/2$, as well as the unique $0^+$ states
with $T = |N-Z|/2 + 2$ are given in Table I.
We have changed the phases of the wave functions that are given in
the technical report \cite{ref1} so that they are consistent with
the convention for coefficients of fractional parentage of 
de-Shalit and Talmi \cite{ref6} and Talmi \cite{ref2}.
The $T = |N-Z|/2+2$ wave functions are obtained with \underline{any} 
isospin conserving interaction, as they are double analogs of corresponding 
states in the Calcium isotopes.

Note that for $^{44}$Ti the dominant parts of the $J=0$ ground state wave
function consist of $s$ and $d$ couplings. 
That is $(J_P, J_N)$ values of $(0, 0)$ and $(2, 2)$ constitute 
about 95\% of wave function. Note that for the $J=0$ $T=1$ ground 
state of $^{46}$Ti the $(4, 4 \, v=4)$ admixture is larger than
that of $(4, 4 \, v=2)$. This is due in part to the fact that in $^{44}$Ca
the $J=4$ seniority $v=4$ state is lower in energy  by about 0.5 MeV than
the $J=4$ $v=2$ state \cite{ref2}. 
This can be understood by noting that the $v=4$ state can be formed by 
two ``$d$ pairs'' and this is energetically favorable over the 
configuration of the $v=2$ stae which must consist of a single ``$d$ pair''. 
%

%
%

%
If we represent the wave function of a given even-even Ti isotope
by the column vector with components $D^{\alpha I}(J_p,J_n v)$,
we then get the following eigenvalue equation for $I=0$ state,
for the Hamiltonian given by Eq.(\ref{angl}),
\begin{eqnarray}
 (n+1) D(J,J) - (n+1) (j^n J j |\} j^{n+1} j)
        \sum_{J'} (j^n J' j |\} j^{n+1} j) D(J',J')
   = \lambda D(J,J)
\end{eqnarray}
The eigenvalue equation for the seperable interaction is
\begin{eqnarray}
 (j^n J j |\} j^{n+1} j)
        \sum_{J'} (j^n J' j |\} j^{n+1} j) D(J',J')
   = \lambda' D(J,J)
\end{eqnarray}
The eigenvalues are obtained by looking at the results for
the isospin version of the interaction as given in Eq.(\ref{meanv}).
We find that $\lambda' = 0$ for $T = T_{min} = |N-Z|/2$
and $\lambda' = 1$ for $T = T_{min} + 2 = |N-Z|/2 + 2$.
Thus we can write the results 
for the eigenfunction $D(JJ)$
in a more compact way.
\begin{eqnarray}
\sum_J D(JJ) (j^n J j |\} j^{n+1} j) &=& 0 {\rm \ \ \ for \ } T=T_{min}
        \nonumber \\
                     &=& 1 {\rm \ \ \ for \ } T=T_{min} + 2
               \label{npdjj}
\end{eqnarray}
with
\begin{eqnarray}
 D(JJ) &=& (j^n J j |\} j^{n+1} j)       \label{djjt2}
\end{eqnarray} 
for $T = T_{min} + 2$.
We can understand Eq.(\ref{djjt2}) better by
noting that the eigenfunctions for $T_{max} = T_{min}+2$ are
given by $D(JJ) = (j^n J j^2 J |\} j^{n+2} 0)$, i.e.,
they are two particle coefficients of fractional parentage.
However it has been shown by Zamick and Devi \cite{ref5}
in the context of the two to one relations between the spectra
of even-even and corresponding even-odd nuclei that 
the two particle cfp is equal to one particle cfp
 $(j^n J j^2 J |\} j^{n+2} 0) = (j^n J j |\} j^{n+1} j)$ and 
hence the one particle cfp above is equal to $D(JJ)$, Eq.(\ref{djjt2}).
The first part of Eq.(\ref{npdjj}) is just the orthogonality
relation between states with $T_{min}$ and $T_{min}+2$ \cite{ref7}.
The second part of Eq.(\ref{npdjj}) is the normalization condition 
for the state of $T_{min} + 2$.

\section{Application - number of $n$-$p$ pairs in the Ti isotopes}

The work in this section can be regarded as an extension of work done
previously by Moya de Guerra et al \cite{ref3}.
We shall also make a comparison with earlier work on pairs 
by Engel et al in which extensive results were obtained for the number 
of $J=0$ pairs using an isovector pairing interaction \cite{ref4}.
In the present work and in Ref.\cite{ref3} we have results for
any interaction.

Using the results of previous sections 
we can easily get an expression for the number of pairs of particles
with total angular momentum $J_A$.
This is identified as the coefficient of
 $E(J_A) = \bra(j^2) J_A |V| (j^2) J_A\ket$ 
in the expression for the total energy Eq.(\ref{mateleh})
between the wave function of Eq.(\ref{psii}).
The contribution of the number of pairs from the neutron-proton
interaction for an $I = 0$ state of Ti is
\begin{eqnarray}
 {\rm number \ of \ pairs}
 &=& 2n \sum_{J_0} \left| \sum D(JJ) (j^{n-1} J_0 j |\} j^n J) (2J+1) (2J_A+1)
        \left\{\matrix{j & j & J \cr J_0 & j & J \cr J_A & J_A & 0}\right\}
       \right|^2   \nonumber  \\
         \vspace*{0.4in}   
 &=& 2n \sum_{J_0} \left| \sum_J D(JJ) (j^{n-1} J_0 j |\} j^n J)
       \sqrt{(2J+1)(2J_A+1)}
        \left\{\matrix{j & j & J \cr j & J_0 & J_A}\right\} \right|^2
              \label{npint}
\end{eqnarray}
We can obtain the number of even pairs by summing the expression in 
Eq.(\ref{npint}) over $J_A$ even by inserting a factor $(1+(-1)^{J_A})/2$;
for odd pairs the factor $(1-(-1)^{J_A})/2$.
Using techniques similar to those in going from Eq.(\ref{melemh}) 
to Eq.(\ref{angl}) we find
\begin{eqnarray}
 {\rm number \ of \ even \ pairs}
    &=& (n-1) + (n+1) \left|\sum_J D(JJ) (j^n J j |\} j^{n+1} j)\right|^2
          \nonumber \\
 {\rm number \ of \ odd \ pairs}
    &=& (n+1) - (n+1) \left|\sum_J D(JJ) (j^n J j |\} j^{n+1} j)\right|^2
              \label{numpair}  
\end{eqnarray}
The above expressions can be greatly simplified by noting the 
relations given by Eq.(\ref{npdjj}).
Thus we find
\begin{eqnarray}
 {\rm number \ of \ even \ } n-p {\rm \ pairs \ }
     &=& n-1 {\rm \ \ for \ } T = T_{min}  \nonumber \\
     &=& 2n {\rm \ \ \ \ \ for \ } T = T_{min}+2      \label{npeo}  \\
 {\rm number \ of \ odd \ } n-p {\rm \ pairs \ }
     &=& n+1 {\rm \ \ for \ } T = T_{min}  \nonumber \\
     &=& 0  {\rm \ \ \ \ \ for \ } T = T_{min}+2   \nonumber
\end{eqnarray}
That there are no odd $n$-$p$ pairs with $T = T_{min}+2$ follows
from the fact that the higher isospin state is the double analog 
of a state in the Ca isotopes, i.e.,
of a system of identical particles that can only have $T=1$ pairs.

We can now obtain a result for the number of pairs with $J_A=0$
(in which case $J_0=j$). From Eq.(\ref{npint}),
\begin{eqnarray}
 {\rm \ number \ of \ } n-p {\rm \ pairs \ } (J_A=0)
    &=& \frac{2n}{(2j+1)^2} \left|\sum_J D(JJ) (j^{n-1} j j |\} j^n J)
            \sqrt{(2J+1)} \right|^2
\end{eqnarray}
The results are, using Eq.(\ref{npdjj}) \cite{ref7},
\begin{eqnarray}
 D(00) &=& \frac{n}{(2j+1)} \sum_J D(JJ) (j^{n-1} j j |\} j^n J)
             \sqrt{(2J+1)}         \label{d00min} 
\end{eqnarray}
\begin{eqnarray}
 {\rm number \ of \ } n-p {\rm \ pairs \ } (T=T_{min},J_A=0)
    &=& \frac{2|D(00)|^2}{n}  
         \label{npairm}
\end{eqnarray}
for $T=T_{min}$, 
i.e., $|D(00)|^2$, $|D(00)|^2/2$, $D(00)|^2/3$ and $|D(00)|^2/4$ 
for $^{44}$Ti, $^{46}$Ti, $^{48}$Ti and $^{50}$Ti respectively.
For $T=T_{min} + 2$, we can show that the number of pairs is given by
\begin{eqnarray}
 {\rm number \ of \ } n-p {\rm \ pairs \ } (T=T_{min}+2,J_A=0)
   &=& 2n |D(0,0)|^2
    = \frac{2n (2j+1-n)}{(2j+1)(n+1)}
         \label{npairm2}
\end{eqnarray}
Furthermore the values of $|D(00)|^2$ are respectively 1/4, 1/10 and 1/28.
So the number of $J_A=0$ pairs are finally 1, 4/5 and 3/7.

Consider the special case $n=2$ which was previously considered
in Ref. \cite{ref7} and \cite{ref3}.
The cfp's become unity (for even $J$) and so do not play role.
We find 
\begin{eqnarray}
 {\rm number \ of \ pairs} 
    &=& 2n \left|\sum_J D(JJ) \sqrt{(2J+1)(2J_A+1)} 
         \left\{\matrix{j & j & J \cr j & j & J_A}\right\}\right|^2
                   \label{npn2}
\end{eqnarray}
This follows from Eqs.(\ref{npint}) and (A2).
But for $n=2$ ($^{44}$Ti) (see Eq.(\ref{talmi})) the Hamiltonian is
\begin{eqnarray}
 \bra[J'J']_0 |H| [JJ]_0\ket/a &=& 2 \delta_{JJ'}
    - 2 \sqrt{(2J+1)(2J'+1)} \left\{\matrix{j & j & J' \cr j & j & J}\right\}
     \label{ti44h}
\end{eqnarray}
From the isospin point of view the eigenvalue is 3 for all three $T=0$ states.
The first term $2 \delta_{JJ'}$ accounts for 2;
thus the eigenvalue of 
 $- 2 \sqrt{(2J+1)(2J'+1)} \left\{\matrix{j & j & J' \cr j & j & J}\right\}$
is +1.
Because of the three fold degeneracy any linear combination of the
three $T=0$ states is an eigenfunction.
Hence 
\begin{eqnarray}
 -2 \sum_J \sqrt{(2J+1)(2J_A+1)}
   \left\{\matrix{j & j & J_A \cr j & j & J}\right\} D(JJ)
     = D(J_A J_A)
\end{eqnarray}
for $T=0$ states.
But the sum on the left is what appears in the expression for the number of 
pairs (Eq.(\ref{npint}) or (\ref{npn2}))
with the restriction that $J_A$ must be even.
Hence we see that for $^{44}$Ti in the single $j$ shell approximation
the number of neutron-proton pairs for any even $J_A$ is
 $2\times 2 \left|\frac{1}{2} D(J_A J_A)\right|^2
     = \left|D(J_A J_A)\right|^2$.
In a concurrent publication \cite{ref7,ref3} this result was obtained 
by diagonalizing a $9j$ symbol.  
Here we show an alternate derivation.
 
For $^{44}$Ti the number of $n$-$p$ pairs with even $J_A$ is the same as 
the number of $n$-$n$ and $p$-$p$ pairs with same $J_A$ -- in each case
the answer is $|D(J_A J_A)|^2$.
In Ref.\cite{ref7} our expressions are compared to the more
special cases of Engel et al.
They obtained the number of $J_A = 0$ pairs for an isovector pairing
interaction and considered only states with $T = T_{min}$ \cite{ref4}.
We have an alternate expression for the number of $J_A=0$ pairs in the Ti 
isotopes which holds for any interaction -   
namely $2|D(00)|^2/n$ for $T=T_{min}$ (Eq.(\ref{npairm}))
and $2n|D(00)|^2$ for $T=T_{min}+2$ (Eq.(\ref{npairm2})).

\section{Closing remarks}

The most obvious use of coefficients of fractional parentage is to
derive relationships for systems of identical particles.
In this work we show that there are some surprising relationships
involving these cfp's for systems of both neutrons and protons.
We did so by considering the Ti isotopes in single $j$ shell calculations.
We exploited the fact that the simplified interaction
 $a + b t(1)\cdot t(2)$ could be evaluated in both isospin space
and angular momentum space.

We found several interesting relations in the text, e.g., 
Eqs.(\ref{npdjj}), (\ref{npeo}), (\ref{d00min}) and (\ref{npairm}).
This leads to the fact that 
for $T=T_{min}$ the number of even pairs is $(n-1)$
and the number of odd pairs is $(n+1)$, 
whilst the number of $J_A=0$ pairs is $2|D(00)|^2/n$.
It is important to get relations for general interactions because,
as shown in Refs.\cite{ref7,ref3}, 
the simple schematic interactions such as $J=0$ $T=1$ pairing
or $J=1$ $T=0$ pairing do not yield good wave functions for systems
of combined neutrons and protons.
We have also applied our results to counting $n$-$p$ pairs of a
given angular momentum in the Ti isotopes and showed how the
expressions could in some case be greatly simpified.
This is of relevance to two nucleon transfer, where the motivation
is to test the relative importance of say $J=0$ $T=1$ pairing 
vs $J=1$ $T=0$ pairing.

But our main conclusion is that there are hidden relationships
to be uncovered even for complex systems involving neutrons and
protons and we hope this work will stimulate others
to look for such relationships.

We thank Martin Redlich from Berkeley for a careful reading of this
manuscript and for his interest.
This work was supported in part by DOE grant DE-FG01-05ER05-02,
in part by DOE grant DE-FG02-96ER-40987, and
in part by Grant 2001-1-11100-005-3 from the Basic Research
Program of the Korea Science and Engineering Foundation.

\appendix

\section{}

Useful relations used in this work (from I. Talmi \cite{ref2}).

\begin{eqnarray}
   & &
 \left\{\matrix{J_1 & J_2 & J \cr J_4 & J_3 & J'}\right\}
   = (-1)^{J_2+J+J_3+J'} \sqrt{(2J+1)(2J'+1)}
      \left\{\matrix{J_1 & J_2 & J \cr J_3 & J_4 & J \cr J' & J' & 0}\right\}
               \\   \vspace{0.4in}   & &
 \sum_{J_{13} J_{24}} (-1)^{J_1+J_4+J_{24}} (-1)^{J_2+J_3+J_{23}}
        (2J_{13}+1) (2J_{24}+1)
     \left\{\matrix{J_1 & J_3 & J_{13} \cr J_2 & J_4 & J_{24} \cr
            J_{12} & J_{34} & J}\right\}
     \left\{\matrix{J_1 & J_4 & J_{14} \cr J_3 & J_2 & J_{23} \cr
            J_{13} & J_{24} & J}\right\}
        \nonumber  \\   & & \hspace{1.1in} 
   = (-1)^{J_3+J_4+J_{34}}
     \left\{\matrix{J_1 & J_4 & J_{14} \cr J_2 & J_3 & J_{23} \cr
            J_{12} & J_{34} & J}\right\}
\end{eqnarray}

cfp recursion formula (from I. Talmi \cite{ref2}).
\begin{eqnarray}
 n (j^{n-1} \alpha_0 J_0 j |\} j^n J) (j^{n-1} \alpha_1 J_1 j |\} j^n J)
   &=& \delta_{\alpha_0 \alpha_1} \delta_{J_1 J_0} 
   + (n-1) \sqrt{(2J_0+1)(2J_1+1)}
   \sum_{J_2} \left\{\matrix{J_2 & j & J_1 \cr J & j & J_0}\right\} 
         \nonumber  \\  & & \hspace{0.3in} \times
      (-1)^{J_0+J_1} (j^{n-2} J_2 j |\} j^{n-1} \alpha_0 J_0)
      (j^{n-2} J_2 j |\} j^{n-1} \alpha_1 J_1)
\end{eqnarray}

Values of some cfp's from de Shalit and Talmi \cite{ref6}
\begin{eqnarray}
 (j^{n-1} j ; j |\} j^n J=0) &=& 1      \nonumber \\
 (j^{n-1} j ; j |\} j^n J v=2) &=& \sqrt{\frac{2(2j+1-n)}{n(2j-1)}} \nonumber \\
 (j^n J=0 ; j |\} j^{n+1} j) &=& \sqrt{\frac{(2j+1-n)}{(n+1)(2j+1)}}
                    \label{eqa5} \\
 (j^n J v=2 ; j |\} j^{n+1} j)
      &=& - \sqrt{\frac{2n (2J+1)}{(n+1)(2j+1)(2j-1)}}
              \nonumber 
\end{eqnarray}

%
%
\begin{table}
\caption{Wave functions of $I=0_1$, $T_{min}$ 
and $I=0$, $T_{min} +2$ states of $^{44}$Ti, $^{46}$Ti and $^{48}$Ti,
represented by column vectors $D(J,Jv)$.}
\begin{tabular}{rcccc}
  \hline 
 $^{44}$Ti & $J_P$  &  $J_N$  &  $I=0$ $T=0$   &   $I=0$ $T=2$  \\
           &   0    &    0    &    0.7608      &   0.5000       \\
           &   2    &    2    &    0.6090      &  --0.3727  \    \\
           &   4    &    4    &    0.2093      &  --0.5000  \    \\
           &   6    &    6    &    0.0812      &  --0.6009  \    \\
  \hline
 $^{46}$Ti & $J_P$  &  $J_N$  &  $I=0$ $T=1$   &   $I=0$ $T=3$  \\
           &   0    &    0    &    0.8224      &   0.3162      \\
           &   2    &    2    &    0.5420      &  --0.4082 \     \\
           &   2    &  2$^*$  &    0.0563      &   0.0          \\
           &   4    &    4    &    0.0861      &  --0.5477 \     \\
           &   4    &  4$^*$  &  --0.1383 \    &   0.0          \\
           &   6    &    6    &  --0.0127 \    &  --0.6583 \     \\
  \hline
 $^{48}$Ti & $J_P$  &  $J_N$  &  $I=0$ $T=2$   &   $I=0$ $T=4$  \\
           &   0    &    0    &    0.9136      &   0.1890       \\
           &   2    &    2    &    0.4058      &  --0.4226 \     \\
           &   4    &    4    &    0.0196      &  --0.5669 \     \\
           &   6    &    6    &  --0.0146 \    &  --0.6814 \      
 \\  \hline
\end{tabular} 

* means $v = 4$.
\end{table}

\end{document}